\documentclass[preprints,article,accept,oneauthor,pdftex,10pt,a4paper]{mdpi} 

\firstpage{1} 
\pubvolume{xx}
\issuenum{1}
\articlenumber{1}
\pubyear{2018}
\copyrightyear{2018}
\history{}

\Title{A classical interpretation of the Scrooge distribution}

\Author{William K. Wootters \orcidA{0000-0001-7117-1366}}

\AuthorNames{William K. Wootters}

\address{%
 \quad Department of Physics, Williams College, Williamstown MA 01267, USA; william.wootters@williams.edu
}

\corres{Correspondence: william.wootters@williams.edu; Tel.: +1-413-597-2156}

\abstract{The Scrooge distribution is a probability distribution over the set of pure states of a 
quantum system.  Specifically, it is the distribution that, upon measurement, 
gives up the {\em least} information about the
identity of the pure state, compared with all other distributions having the same density
matrix.  The Scrooge distribution has normally been regarded as a purely quantum mechanical
concept, with no natural classical interpretation.  In this paper we offer a classical interpretation
of the Scrooge distribution viewed as a probability distribution over the probability simplex.  
We begin by considering a real-amplitude version of the Scrooge distribution, for which we find
that there is a non-trivial but natural classical interpretation.  The transition to the complex-amplitude
case requires a step that is not particularly natural but that may shed light on the relation between
quantum mechanics and classical probability theory. }


\begin{document}


\section{Introduction}

In the early days of quantum information theory, the term ``quantum communication'' would typically
have been understood to refer to 
the transmission of {\em classical} information via quantum mechanical signals.  
Such communication can be done in a sophisticated way, with the receiver making
joint measurements on several successive signal particles \cite{Schumacher1,Schumacher2}.  Or it can be done in a 
relatively straightforward way, with the receiver performing a separate measurement on each
individual signal particle.  In both cases, but especially in the latter case, a particularly interesting quantity, given an ensemble of quantum states to be used as an alphabet, is the ensemble's {\em accessible information}.
This is the maximum amount of information one can obtain about the identity of the state, on average, by
making a measurement on the system described by the specified ensemble. 
{The average here is over the outcomes of the measurement, and the maximization is
over all possible measurements.  In general,
accessible information can be defined for ensembles consisting of pure and mixed states, but in 
this paper we consider only pure-state ensembles.}

Given an ensemble $\{(|\psi_j\rangle, p_j)\}$ of pure quantum states with their probabilities, there is a unique density matrix associated with
the ensemble.  But for any given {density matrix
representing more than a single pure state}, there are infinitely many
ensembles described by that density matrix; let us call them $\rho$-ensembles.  Thus it is natural to ask this question: for a given density matrix $\rho$, what pure-state $\rho$-ensemble
has the greatest value of the accessible information, and what pure-state
$\rho$-ensemble has the least value of the accessible information?  The former question 
was answered by an early (1973) result 
in quantum information
theory \cite{Holevo}: the pure-state $\rho$-ensemble with the 
greatest accessible information is the one consisting of the eigenstates of $\rho$, with weights
given by the eigenvalues of $\rho$.  The latter question was answered in a 1994 paper \cite{JRW}, in which
the $\rho$-ensemble minimizing the accessible information was called the Scrooge ensemble, or Scrooge distribution, since it is the ensemble most stingy with its 
information.  

To see a simple example, consider a spin-1/2 particle whose density matrix $\rho$ has as its
eigenvectors the states $\mid\uparrow\rangle$ and $\mid\downarrow\rangle$, with
eigenvalues $\lambda_\uparrow$ and $\lambda_\downarrow$.  The eigenstate ensemble for $\rho$, that is, the $\rho$-ensemble from which one can extract the most information, is the two-state ensemble consisting of the state
$\mid\uparrow\rangle$ with probability $\lambda_\uparrow$
and the state $\mid\downarrow\rangle$ with probability $\lambda_\downarrow$.  
The optimal measurement in this case---the measurement that provides the most information---is the up-down measurement, and the amount of information it provides is equal to the von Neumann
entropy of the density matrix:
\begin{equation}
I = S(\rho) = -(\lambda_\uparrow \ln \lambda_\uparrow + \lambda_\downarrow \ln \lambda_\downarrow)
\end{equation}
(In this paper we use the natural logarithm in all information-theoretic quantities.)  

On the other hand, the Scrooge ensemble for this density matrix is represented by a continuous
probability distribution over the whole surface of the Bloch sphere.  If $\lambda_\uparrow$ is larger than
$\lambda_\downarrow$, then this continuous distribution is weighted more heavily toward the 
top of the sphere.  We can write the Scrooge distribution explicitly in terms of the variable
$x = (1+\cos\theta)/2$, where $\theta$ is the angle measured from the north pole:  
\begin{equation}
\sigma(x) = \frac{2}{\lambda_\uparrow \lambda_\downarrow}\cdot\frac{1}{\big(\frac{x}{\lambda_\uparrow} + \frac{1-x}{\lambda_\downarrow}\big)^3}.
\end{equation}
The probability density $\sigma(x)$ is normalized in the sense that 
$\int_0^1 \sigma(x) dx = 1$.  
(The distribution is uniform over
the azimuthal angle $\phi$.)  Again, this is the ensemble of pure states from which one can extract the least information about the identity
of the pure state, among all ensembles having the density matrix $\rho$.  Somewhat remarkably, the average amount of information one gains 
by measuring this particular ensemble is entirely independent of the choice
of measurement, as long as the measurement is complete---that is, as long as each
outcome is associated with a definite pure state.  This amount of information comes out to be
a quantity called the subentropy $Q$ of the density matrix:
\begin{equation}  \label{subentropy2}
I = Q(\rho) = - \frac{\lambda_\uparrow^2 \ln \lambda_\uparrow - \lambda_\downarrow^2 \ln \lambda_\downarrow}{\lambda_\uparrow - \lambda_\downarrow}.
\end{equation}
We will give more general expressions for both the Scrooge ensemble and the subentropy in Section 2 below.  

In recent years, the Scrooge distribution has made other appearances in the physics literature.  
Of particular interest is the fact that this
distribution has 
emerged 
from an entirely different line of investigation,
in which the system under consideration is entangled with a large environment and the whole system
is in a pure state.  In that case, if one looks at the {\em conditional} pure states
of the original system, relative to the elements of an orthogonal basis of the environment, 
one typically finds that these conditional states are distributed by a Scrooge distribution \cite{Goldstein1,Tumulka,Reimann,Goldstein2}.
In this context the distribution is usually called a GAP measure (for {\bf G}aussian {\bf a}djusted
{\bf p}rojected measure, the three 
adjectives corresponding to the three steps by which the measure can be 
constructed).   On another front, the Scrooge distribution has been {used} to address the difficult problem of bounding the {\em locally}
accessible information when there is more than one receiver \cite{SenDe}. 

Meanwhile, the concept of subentropy, which originally arose (though without a name) in connection with 
the outcome-entropy of random measurements \cite{Wootters,Jones}, has appeared not only in problems concerning the acquisition
of classical information \cite{Jacobs1,Jacobs2,DallArno}, but also in the 
quantification of entanglement \cite{Mintert} and the study of quantum coherence \cite{Cheng,Zhang1,Zhang2,Zhang3}.  Many detailed properties of subentropy have now been worked out,
especially concerning its relation to the Shannon entropy \cite{Nichols, Jozsa1,Jozsa2,Jozsa3,Datta}.

Though it is possible to devise a strictly classical 
situation in which subentropy arises \cite{Jozsa2}, the Scrooge distribution has generally been regarded as 
a purely quantum mechanical concept.  It is, after all, a probability
distribution over pure quantum states.  
The aim of this paper is to provide a classical interpretation of the Scrooge distribution,
and in this way to provide a new window into the relation between quantum mechanics
and classical probability theory.  

We find that it is much easier to make the connection
if we begin by considering not the standard Scrooge distribution, but rather the analogous distribution
one obtains for the case of quantum theory with {\em real} amplitudes.  In that case,
the dimension of the set of pure states is the same as the dimension of the associated
probability simplex, and we find that there is a fairly natural distribution within classical probability
theory that is essentially identical to the real-amplitude version of the Scrooge distribution.  
This distribution arises as the solution to a certain classical communication
problem that we describe in Section 4.  

With this interpretation of the real-amplitude Scrooge distribution in hand, we ask how the
classical communication scenario might be modified so as to arrive at the original Scrooge ensemble
for standard, complex-amplitude quantum theory.  As we will see, the necessary modification is not particularly
natural, but it is simple.

Thus we begin in Sections 2 and 3 by reviewing the derivation of the Scrooge distribution
and by working out the analogous distribution for the case of real amplitudes.  
Then in Section 4 we set up and analyze the classical communication problem that, as
we show in Section 5, gives rise to a distribution equivalent to the real-amplitude Scrooge distribution.
In Section 6 we modify the classical communication scenario to
produce the standard, complex-amplitude Scrooge distribution.  Finally, we summarize and discuss
our results in Section 7.

\section{The Scrooge distribution}

There are several ways in which one can generate the Scrooge distribution.  In this section we
review the main steps of the derivation given in Ref.~\cite{JRW}, which applies to 
a Hilbert space of finite dimension.  (The distribution can also be defined for
an infinite-dimensional Hilbert space \cite{Goldstein1,Tumulka,Reimann,Goldstein2}.)  
We begin by setting up the problem.

We imagine the following scenario.  One participant, Alice, prepares a quantum system having an 
$n$-dimensional Hilbert space in 
a pure state $|x\rangle$ and sends it to Bob.  Bob then tries to gain information about the identity of
this pure state.  Initially, Bob's state of
knowledge is represented by a probability density $\sigma(x)$ over the 
set of pure states.  (The symbol $x$ represents a multi-dimensional parameterization of the
set of pure states.)  Bob makes a measurement on the system and thereby gains
information.  The amount of information he gains may depend on the outcome he obtains, so we
consider the {\em average} amount of information he gains about $x$, the 
average being over all outcomes.  

The standard quantification
of Bob's average gain in information is the Shannon mutual information between
the identity of the pure state and the outcome of the measurement.  We can express this
mutual information in terms of two probability functions: (i) the probability $p(j|x)$ of the outcome $j$ when the state is
$|x\rangle$, and (ii) the overall probability $p(j) = \int \sigma(x) p(j|x) dx$ of the outcome $j$ averaged over the whole 
ensemble.  In terms of these functions, the mutual information is
\begin{equation}
I = - \sum_j p(j) \ln p(j) + \int \sigma(x) \Big[ \sum_j p(j|x) \ln p(j|x) \Big] dx.
\end{equation}
The {\em accessible information} of the ensemble defined by $\sigma(x)$ is the maximum
value of the mutual information $I$, where the maximum is taken over all possible measurements.  

Again, for a given density matrix $\rho$, the Scrooge distribution is 
defined to be the {pure-state} $\rho$-ensemble with
the least value of the accessible information.  One can obtain
the Scrooge distribution via the following algorithm \cite{JRW}.

We start by recalling the concept of ``$\rho$ distortion.''
Consider for now a finite ensemble $\{(|\psi_i\rangle, p_i)\}$ of pure states, $i = 1, \ldots, m$, whose density matrix is the completely mixed state:
\begin{equation}
\sum_{i=1}^m p_i |\psi_i\rangle\langle \psi_i| = \frac{1}{n}I.
\end{equation}
Let $|\tilde{\psi}_i\rangle$ be the subnormalized state vectors
$|\tilde{\psi}_i\rangle = \sqrt{p_i}|\psi_i\rangle$, so that
\begin{equation}
\sum_{i=1}^m |\tilde{\psi}_i\rangle\langle \tilde{\psi}_i | = \frac{1}{n}I.
\end{equation}
Under $\rho$ distortion, each vector $|\tilde{\psi}\rangle$ is mapped to 
another subnormalized vector $|\tilde{\phi}\rangle$ defined by
\begin{equation}
|\tilde{\phi}_i\rangle = \sqrt{n\rho}|\tilde{\psi}_i\rangle.
\end{equation}
Note that the density matrix formed by the $|\tilde{\phi}_i\rangle$'s is $\rho$:
\begin{equation}
\sum_{i=1}^m |\tilde{\phi}_i\rangle\langle \tilde{\phi}_i | = \sqrt{n\rho}\left(\frac{1}{n}I\right) \sqrt{n\rho} = \rho.
\end{equation}
In terms of normalized vectors, the new ensemble is $\{(|\phi_i\rangle, q_i)\}$, with the 
new probabilities $q_i$ equal to 
\begin{equation} \label{newprob}
q_i = \langle \tilde{\phi}_i|\tilde{\phi}_i \rangle = n p_i \langle \psi_i |\rho | \psi_i \rangle.
\end{equation} 
In this way, any ensemble having the completely mixed density matrix can be mapped
to a ``$\rho$ distorted'' ensemble with density matrix $\rho$.  

The Scrooge ensemble is a continuous ensemble, not a discrete one, but the concept of
$\rho$ distortion can be immediately extended to the continuous case, and the Scrooge
distribution can be easily characterized in those terms:  it is the $\rho$ distortion of the {\em uniform} distribution over the unit
sphere in Hilbert space.  {(The uniform distribution is 
the unique probability distribution over the set of pure states that is invariant under all unitary
transformations.)}

Let us see how the $\rho$ distortion works out in this case.  
First, for the uniform distribution, it will be convenient to label the parameters of the pure
states by $y$ instead of $x$, so that we can reserve $x$ for the Scrooge distribution.  Let $\tau(y)$ be the probability density over $y$ that represents the uniform distribution over the unit sphere.  (A particular parameterization will be specified shortly.)  In terms of normalized states, a $\rho$ distortion
maps each pure state $|y\rangle$ into the pure state
$|x\rangle$ defined by
\begin{equation} \label{contdist}
|x\rangle = \frac{\sqrt{\rho}|y\rangle}{\sqrt{\langle y | \rho | y\rangle}}.
\end{equation}
This mapping defines $x$ as a function of $y$: $x = f(y)$.  
(We write $f$ explicitly below.)
The resulting probability density over $x$ is obtained from the continuous version of 
Eq.~(\ref{newprob}).
\begin{equation} \label{sigmay}
\sigma(x) = n \tau(y) \langle y |\rho |y \rangle {\mathcal J}(y/x).
\end{equation}
Here ${\mathcal J}(y/x)$ is the Jacobian of the $y$ variables with respect to the $x$ 
variables.  On the right-hand side of Eq.~(\ref{sigmay}), we interpret each $y$ as
$f^{-1}(x)$, so that we get an expression depending only on $x$.  

To get an explicit expression for the Scrooge distribution---that is, an explicit
expression for the probability density $\sigma(x)$---we need to choose a specific
set of parameters labeling the pure states.  We will choose the same set of parameters to label
both the uniform distribution (where we call the parameters $y$) and the Scrooge distribution (where
we call the parameters $x$).  We define our 
parameters relative to a set of normalized eigenstates $|e_j\rangle$ 
of the density matrix $\rho$.  A general pure state $|x\rangle$ can be written as 
\begin{equation}
|x\rangle = \sum_{j=1}^n a_j e^{-i\theta_j} |e_j\rangle,
\end{equation}
where each $a_j$ is a non-negative real number and each phase $\theta_j$ runs from zero to $2\pi$.  
For definiteness, employing the freedom to choose an overall phase, we define $\theta_n$ to be zero.
We take $x$ (or $y$) to consist of the following parameters: the squared amplitudes $x_j = a_j^2$ for $j = 1, \ldots, n-1$, and 
the phases $\theta_j$ for $j =1, \ldots, n-1$.  This set of $2n - 2$ parameters uniquely identifies
any pure state.  Later we will also use the symbol $x_n = 1 - x_1 - \cdots - x_{n-1}$.  
Note that the $x_j$'s are the probabilities of the outcomes of a particular orthogonal measurement, 
whose outcomes are associated with the eigenstates of $\rho$.

In terms of these parameters, the uniform distribution over the unit sphere takes a particularly
simple form: it is the product of a uniform distribution over the phases and a uniform distribution
over the $(n-1)$-dimensional
probability simplex whose points are labeled by $\{x_1, \ldots, x_{n-1}\}$ \cite{Sykora}.  The 
Scrooge distribution will likewise be a product and will be uniform over the phases but will typically have a certain bias over
the probability simplex.  Because the phases are always independent and uniformly distributed 
in the cases we are considering,
we will omit the phases in our expressions for the distributions, writing the probability densities
as functions of $\{x_1, \ldots, x_{n-1}\}$ (or $\{y_1, \ldots, y_{n-1}\}$).  

Our aim now is to find explicit expressions for each of the factors appearing on the
right-hand side of 
Eq.~(\ref{sigmay}).  Since the uniform distribution over the unit sphere induces a uniform distribution over the probability simplex, 
the corresponding probability density $\tau(y)$ is a constant function, the value of the constant
being $(n-1)!$ as required by normalization:
\begin{equation}
(n-1)! \int_0^1 \int_0^{1-y_1} \cdots \int_0^{1-y_1 - \cdots - y_{n-2}} dy_{n-1} \cdots dy_2 dy_{1} = 1.
\end{equation}
The function $f(y)$ defined by the $\rho$-distortion map, Eq.~(\ref{contdist}), is
given by
\begin{equation} \label{f}
x_j = \frac{\lambda_j y_j}{\lambda_1 y_1 + \cdots + \lambda_n y_n}, \hspace{4mm} j = 1, \ldots, n-1,
\end{equation}
where the $\lambda_j$'s are the eigenvalues of the density matrix $\rho$.  One finds
that the inverse map is
\begin{equation} \label{inverse}
y_j = \frac{x_j/\lambda_j}{x_1/\lambda_1 + \cdots + x_n/\lambda_n},
\end{equation}
and the Jacobian is
\begin{equation} \label{Jac}
{\mathcal J}(y/x) = \frac{1}{\lambda_1 \cdots \lambda_n}\cdot \frac{1}{\big( \frac{x_1}{\lambda_1} + \cdots + \frac{x_n}{\lambda_n}\big)^n}.
\end{equation}
Meanwhile the factor $\langle y |\rho| y \rangle$ can be written as
\begin{equation} \label{expect}
\langle y |\rho| y \rangle = \lambda_1 y_1 + \cdots + \lambda_n y_n = \frac{1}{\frac{x_1}{\lambda_1} + \cdots + \frac{x_n}{\lambda_n}}.
\end{equation}
Substituting the expressions from Eqs.~(\ref{Jac}) and (\ref{expect}) into Eq.~(\ref{sigmay}), we finally arrive at the probability 
density defining the Scrooge distribution:
\begin{equation} \label{Scrooge}
\sigma(x) = \frac{n!}{\lambda_1 \cdots \lambda_n} \cdot \frac{1}{\big( \frac{x_1}{\lambda_1} + \cdots + \frac{x_n}{\lambda_n}\big)^{n+1}}.
\end{equation}
This probability density is normalized in the sense that the integral over the probability 
simplex is unity:
\begin{equation} \label{normalize}
\int_0^1 \int_0^{1-x_1} \cdots \int_0^{1-x_1 - \cdots - x_{n-2}} \sigma(x) dx_{n-1} \cdots dx_2 dx_{1} = 1.
\end{equation}

Now, how do we know that the distribution given by Eq.~(\ref{Scrooge}) minimizes the
accessible information?  First, one can show that for this distribution, the mutual information $I$ is independent
of the choice of measurement, as long as the measurement is complete \cite{JRW}.  So one can compute the value of the accessible information by considering any such measurement, and the easiest
one to consider is the orthogonal measurement along the eigenstates.  The result is
\begin{equation} \label{subentropy}
\hbox{accessible information} = -\sum_{k=1}^n \left( \prod_{l \ne k} \frac{\lambda_k}{\lambda_k - \lambda_l} \right) \lambda_k \ln \lambda_k,
\end{equation}
which defines the subentropy $Q$.  One can also show that for {\em any} $\rho$-ensemble,
the {\em average} mutual information over all complete orthogonal measurements is equal to 
$Q$, which implies that $Q$ is always a lower bound on the accessible information.  Since
the Scrooge distribution {\em achieves} the value $Q$, it achieves the minimum
possible accessible information among all $\rho$-ensembles.


\section{The real-amplitude analog of the Scrooge distribution}
Though our own world is described by standard quantum theory with complex amplitudes,
we can also {consider} an analogous, hypothetical theory with real amplitudes.
A pure state in the real-amplitude theory 
is represented by a real unit vector, and a 
density matrix is represented by a symmetric real matrix with non-negative eigenvalues and
unit trace.  Time evolution in this theory is generated by an antisymmetric real operator in
place of the antihermitian operator $iH$.

The question considered in the preceding section can also be asked in the real-amplitude
theory.  Given a density matrix $\rho$, we ask what $\rho$-ensemble has the smallest value
of the accessible information.  It turns out that essentially all of the methods used in the
preceding section continue to work in the real case.  Again one begins with the uniform
distribution over the unit sphere of pure states, and again one obtains the Scrooge
ensemble (in this case the real-amplitude Scrooge ensemble) via $\rho$ distortion.  
The arguments leading to the conclusion that the
ensemble produced in this way minimizes the accessible information work just as
well in the real-amplitude case as in the complex-amplitude case.  

The one essential difference between the two cases lies in the form of initial probability density
$\tau(y)$ associated with the uniform distribution over the unit sphere in Hilbert space.
Whereas in the complex case the induced distribution over the probability simplex is
uniform, in the real case, the induced distribution over the probability simplex is more heavily weighted
toward the edges and corners.  

We can see an example by considering the case $n=2$.  Instead of starting with a uniform distribution over the 
surface of the Bloch sphere, one starts with a uniform distribution over the unit circle in a two-dimensional
real vector space.  Let $\gamma$ be the angle around this circle measured from some chosen axis. 
(Once a density matrix has been specified, we will take this axis to be along one of the eigenstates
of the density matrix.)  
Then $\gamma$ is initially uniformly distributed.  The parameter analogous to $y_1$ of the
preceding section is $y = \sin^2\gamma$.  Note that $y$ runs from 0 to 1 as $\gamma$ runs from 
0 to $\pi/2$.  The initial probability density $\tau_r(y)$ is therefore obtained
from 
\begin{equation}
\tau_r(y)dy = (2/\pi)d\gamma,
\end{equation}
which leads to 
\begin{equation}
\tau_r(y) = \frac{1}{\pi}\cdot \frac{1}{\sqrt{y(1-y)}}.
\end{equation}
(The subscript $r$ is for ``real.'')  
This is in contrast to the function $\tau(y) = 1$ that would apply in the 
complex-amplitude case.  We see that in the real case, $\tau_r(y)$ is largest
around $y=0$ and $y=1$.  

In $n$ dimensions, we take as our parameters specifying a pure state (i) the 
first $n-1$ probabilities $y_j$, $j=1,\ldots, n-1$, of the outcomes of a certain orthogonal measurement (which we will choose
to be the measurement along the eigenvectors of the given density matrix), 
and (ii) a set of discrete phase parameters (each of them taking the values $\pm 1$), which will always be independently and uniformly distributed
and therefore suppressed in our expressions for the probability densities.  

For the uniform distribution over the unit sphere in the $n$-dimensional real Hilbert space,
one can show that the induced distribution over the parameters $(y_1, \ldots, y_{n-1})$
is given by \cite{W1}
\begin{equation} \label{newtau}
\tau_r(y) = \frac{\Gamma(n/2)}{\pi^{n/2}}\cdot \frac{1}{\sqrt{y_1 \cdots y_n}},
\end{equation}
where $y_n = 1 - y_1 - \cdots - y_{n-1}$.  This probability density is normalized over
the probability simplex as
in Eq.~(\ref{normalize}):
\begin{equation}
\int_0^1 \int_0^{1-y_1} \cdots \int_0^{1-y_1 - \cdots - y_{n-2}} \tau_r(y) dy_{n-1} \cdots dy_2 dy_{1} = 1.
\end{equation}

The general expression for $\sigma(x)$ given in Eq.~(\ref{sigmay}) remains valid in the real case, as do
the equations (\ref{inverse}), (\ref{Jac}), and (\ref{expect}) for the various factors in 
Eq.~(\ref{sigmay}).  Again, the one difference is in $\tau_r(y)$, for which we now use 
Eq.~(\ref{newtau}).  Combining these ingredients, we arrive at our expression for the real-amplitude
Scrooge ensemble:
\begin{equation} \label{realScrooge}
\sigma_r(x) = \frac{n \Gamma(n/2)}{\pi^{n/2} \sqrt{\lambda_1 \cdots \lambda_n}\sqrt{x_1 \cdots x_n} \big(\frac{x_1}{\lambda_1} + \cdots + \frac{x_n}{\lambda_n}\big)^{\frac{n}{2} + 1}},
\end{equation}
where, as before, the $\lambda_j$'s are the eigenvalues of the density matrix whose 
Scrooge distribution we are computing.

Though Eq.~(\ref{realScrooge}) has been derived as a distribution over the set of pure states
in real-amplitude quantum theory, it reads as a probability distribution over the 
\hbox{$(n-1)$}-dimensional probability simplex for a classical random variable with
$n$ possible values.  One can therefore at least imagine that there
might be a classical scenario in which this distribution is natural.  In the following section
we identify such a scenario.


\section{Communicating with dice}
Ref.~\cite{W1} imagines the following classical communication scenario.  
Alice is trying to convey to Bob the location of a point in an $(n-1)$-dimensional
probability simplex.  
To do this, she constructs a weighted $n$-sided die that, for Bob, has the probabilities corresponding
to the point Alice is trying to convey.  She then sends the die to Bob, who rolls the
die many times in order to estimate the probabilities of the various possible outcomes.  
But the information transmission is limited in that Bob is allowed only a fixed number of rolls---let us
call this number $N$.  (Perhaps the die automatically self-destructs after $N$ rolls.)  So Bob will always 
have an imperfect estimate of the probabilities Alice is trying to convey.  
Alice and Bob are allowed
to choose in advance a discrete set of points in the probability simplex---these are the 
points representing the set of signals Alice might try to send---and they choose this set
of points, along with their {\em a priori} weights, 
so as to maximize the mutual information between the identity of the point being conveyed
and the result of Bob's rolls of the die.  The main result of that paper is that in the limit of
large $N$, the optimal distribution of points in the probability simplex approximates the
continuous distribution over the simplex expressed by the following probability 
density:
\begin{equation} \label{newtauagain}
\hat{\tau}(y) = \frac{\Gamma(n/2)}{\pi^{n/2}}\cdot \frac{1}{\sqrt{y_1 \cdots y_n}},
\end{equation}
where the $y_j$'s are the probabilities.  (We use a hat in our labels of 
probability densities that arise in a classical context.)  This result is interesting because it is the
same probability density as the one induced by the uniform distribution over the
unit sphere in real Hilbert space (Eq.~(\ref{newtau}) above).  Thus, 
in a world based on real-amplitude quantum theory, as opposed to 
the complex-amplitude theory, there is a sense in which one could say that nature optimizes
the transfer of information.  

That paper---and closely related papers \cite{W2,W3}---deal only with the 
uniform distribution over the unit sphere, not with non-trivial
Scrooge distributions.  In the present section we consider a modification of the above
communication scenario, and in the next section we show this modified 
scheme yields the real-amplitude Scrooge distribution.

A natural way to generalize the above communication scheme is this: let the allowed number $N$ of rolls of the die
vary from one die to another.  (That is, some dice last longer than others before they self-destruct.)  
And once we allow
$N$ to vary, it makes sense to let $N$ itself be another random variable that conveys information.
We are thus led to consider the following scenario.  

Alice is trying to convey to Bob an ordered $n$-tuple of non-negative real numbers
$(M_1, \ldots, M_n)$.  (Alice and Bob will agree in advance on a specific set of such ordered
$n$-tuples, any one of which Alice might try to convey.)  Let us refer to such an
$n$-tuple as a ``signal.'' In order to convey her signal, Alice will send Bob
an $n$-sided die that Bob then begins to roll, over and over, keeping track of the number of 
times each outcome occurs.  Let $N_j$ be the number of times the outcome $j$ occurs.  
At some point, the die self-destructs.  Alice has constructed both 
the weighting of the die
and the self-destruction mechanism so that the {\em average} value of $N_j$ is $M_j$.  

But both the rolling of the die and its duration are probabilistic, and Alice
cannot completely control either the individual numbers $N_j$ or their sum.  For
any given signal $(M_1, \ldots, M_n)$, we assume that
each $N_j$ is distributed independently according to a Poisson distribution with mean value $M_j$:
\begin{equation}
P(N_1, \ldots, N_n|M_1, \ldots, M_n) = \prod_{j=1}^n e^{-M_j} \frac{M_j^{N_j}}{N_j!}.
\end{equation}
This is equivalent
to assuming that the {\em total} number $N$ of rolls of the die is Poisson distributed with
mean value $M = M_1 + \cdots + M_n$, and that for a given total
number of rolls, the numbers of occurrences of the individual outcomes are distributed according
to a multinomial distribution with weights $M_j/M$.  That is, we are assuming the usual statistics
for rolling a die, together with a Poisson
distribution for the total number of rolls.  (Another model we could have used is to have Alice send Bob a radioactive sample
that can decay in $n$ ways, which Bob is allowed to observe with detectors for a fixed amount
of time.)

To make the problem interesting, and to keep Alice from being able to send Bob an arbitrarily 
large 
amount of information in a single die, we place limits on the sizes of $M_1, \ldots, M_n$.
We do this by imposing, for each $j$, an upper bound ${\mathcal M}_j$ (script $M$) on the expectation value of the number of times the $j$
outcome occurs.  This expectation value is an average over all the possible signals that
Alice might send.  

We also need to say in what sense Alice and Bob are optimizing their communication.  There are a number of reasonable options for doing this---e.g., we could say they maximize the mutual information, or minimize the probability of error {for a fixed number of signals}---but
it is likely that many of these formulations will be essentially equivalent when the values
${\mathcal M}_j$ become very large.  Here we take a simple, informal approach:
we say that, in order to make the various signals distinguishable from each other, Alice and Bob choose their $n$-tuples $(M_1, \ldots, M_n)$ so that
neighboring signals, say, $(M_1, \ldots, M_n)$ and \hbox{$(M_1 +\Delta M_1, \ldots, M_n +\Delta M_n)$},
are at least a certain distance from each other, where we 
use the Fisher information metric to measure distance.  Specifically, we require
the Fisher information distance between the probability distributions
$P(N_1, \ldots, N_n|M_1, \ldots, M_n)$ and
\hbox{$P(N_1, \ldots, N_n|M_1+\Delta M_1, \ldots, M_n+\Delta M_1)$} to
be greater than or equal to a specified value $d_{min}$.  (Or, equivalently for
small $\Delta M_j/M_j$, we require the Kullback-Leibler divergence to be at least $(1/2)d_{min}^2$.)  For the Poisson distribution and for small values of the ratios $\Delta M_j/M_j$, this condition works out to be
\begin{equation} \label{condition}
\sum_{j=1}^n \frac{(\Delta M_j)^2}{M_j} \ge d_{min}^2.
\end{equation}
(For our purposes the exact value of $d_{min}$ will not be important.)  
We also assume that the various signals have equal {\em a priori}
probability.  This is a natural choice if one wants to convey as much information as possible.  
Under these assumptions, Alice and Bob's aim is to maximize the number of distinct signals.

The analysis will be much simpler if we parameterize each die not by 
$(M_1, \ldots, M_n)$, but rather by the variables
\begin{equation}
\alpha_j = \sqrt{M_j}, \hspace{3mm} j=1, \ldots, n.
\end{equation}
Then for neighboring signals we can write 
\begin{equation}
\Delta \alpha_j =\frac{1}{2\sqrt{M_j}}\Delta M_j,
\end{equation}
so that 
the 
condition in Eq.~(\ref{condition}) becomes
\begin{equation}
\sum_{j=1}^n (\Delta\alpha_j)^2 \ge \frac{1}{4}\, d_{min}^2.
\end{equation} 
That is, in the space parameterized by $\vec{\alpha} = (\alpha_1, \ldots, \alpha_n)$, we want the points
representing Alice's signals to be evenly separated from each other.  Thus Alice's 
signals will be roughly uniformly distributed over some region of $\vec{\alpha}$-space: she wants to 
pack in as many signals as possible without exceeding the bounds
${\mathcal M}_j$ on the expectation values of the $N_j$'s.  In what follows, we 
approximate this discrete but roughly uniform distribution of the values of $\vec{\alpha}$
by a continuous probability distribution.  The probability density 
is zero outside the region where Alice's possible signals
lie; inside that region, it has the constant value $1/V$, where $V$ is the volume of the 
region.  

The communication problem then becomes a straightforward geometry problem: within
the ``positive'' section of $\vec{\alpha}$-space (that is, the section in which
each $\alpha_j$ is non-negative), our aim is to find the region ${\mathcal R}$ of largest volume that 
satisfies the 
constraints 
\begin{equation} \label{geocond}
\frac{1}{V_{\mathcal R}}\int_{\mathcal R} \alpha_j^2 \, d\vec{\alpha} = {\mathcal M}_j, \hspace{3mm} j=1,\ldots, n,
\end{equation}
where $V_{\mathcal R}$ is the volume of ${\mathcal R}$.
It is because Alice's signals have a fixed packing density within ${\mathcal R}$ that we are maximizing the volume:
the larger the volume, the 
more signals Alice has at her disposal.  

It is not hard to see that the solution to this geometry problem is to make the region
${\mathcal R}$ the positive section of a certain ellipsoid centered at the origin.  To see this,
we first rewrite the conditions (\ref{geocond}) as
\begin{equation}
\int_{\mathcal R} \alpha_j^2 \, d\vec{\alpha} = {\mathcal M}_j \int_{\mathcal R} d\vec{\alpha},\hspace{3mm} j=1,\ldots, n.
\vspace{1mm}
\end{equation}
Now let $\beta_j = \alpha_j\Big/\sqrt{\mathcal M_j}$.  In terms of the $\beta_j$'s, the above conditions become
\begin{equation}  \label{othergeo}
\int_{\mathcal R'} \beta_j^2 \, d\vec{\beta} = \int_{\mathcal R'} d\vec{\beta}, \hspace{3mm} j=1,\ldots, n,
\end{equation}
where ${\mathcal R'}$ is the region of $\vec{\beta}$-space corresponding to the region
${\mathcal R}$ of $\vec{\alpha}$-space.  In particular, the equation obtained by summing
these $n$ conditions must also be true:
\begin{equation}
\int_{\mathcal R'} \beta^2 \, d\vec{\beta} = n \int_{\mathcal R'} d\vec{\beta},
\end{equation}
where $\beta^2 = \beta_1^2 + \cdots + \beta_n^2$.
That is, the average squared distance from the origin over the region ${\mathcal R'}$
must be equal to $n$.  
The maximum-volume region ${\mathcal R'}$ satisfying this one condition is the positive
section of a sphere, and one can work out that the radius of the sphere must be $\sqrt{n+2}$.  But that region also satisfies all the conditions (\ref{othergeo}).
So that same region is the maximum-volume region satisfying those conditions as well.  Transforming
back to the variables $\alpha_j$, we see that the maximum-volume region satisfying
the conditions (\ref{geocond}) is the positive section of an ellipsoid, with semi-axis
lengths
\begin{equation}
\alpha_j^{max} = \sqrt{(n+2){\mathcal M}_j}.
\end{equation}

Thus, the strategy Alice and Bob adopt is to choose a set of closely packed signals,
with some minimum separation in $\vec{\alpha}$-space, that occupy the positive section of
an ellipsoid centered at the origin.  Again, in this paper
we are treating this discrete but roughly uniform distribution of signals as if it were actually
uniform.  This approximation becomes more and more reasonable as the values of the
${\mathcal M}_j$'s increase.


\section{A distribution over the probability simplex}
So far, we have not made any connection between our communication problem and
the real-amplitude Scrooge distribution.  We do this now by seeing how the
uniform distribution over the ellipsoid in $\vec{\alpha}$-space induces a certain probability
distribution over the $(n-1)$-dimensional probability simplex for Alice's $n$-sided
dice.  We define this probability distribution as follows.

Let us imagine many rounds of communication from Alice to Bob: she has sent him many dice,
for which the expected numbers of occurrences of the various outcomes, $(M_1, \ldots, M_n)$, cover
a representative range of values: the corresponding values of $\vec{\alpha}$ are distributed fairly uniformly over the region ${\mathcal R}$
in $\vec{\alpha}$-space.  Bob has rolled each of these dice as many times as the die could
be rolled.
Now consider a small region
of the probability simplex, say, the region ${\mathcal S}(x,\Delta x)$
for which the probability of the $j$th outcome lies between $x_j$ and $x_j + \Delta x_j$ for
$j = 1, \ldots, n-1$.  Some of the dice Alice has sent to Bob have probabilities lying in this region.  
The weight we want to attach
to the region ${\mathcal S}(x, \Delta x)$ is, roughly speaking, the fraction of the total number
of rolls that came from dice in this region.  Note that for a die at location 
$\vec{\alpha}$, the expectation
value of the number of times it will be rolled 
is $\alpha^2 = \alpha_1^2 + \cdots + \alpha_n^2$.  So we multiply the density 
of signals by the factor $\alpha^2$ to get the ``density of rolls.''
These considerations lead us to the following definition of the 
weight $\hat{\sigma}(x) dx_1 \cdots dx_{n-1}$ we assign 
to the infinitesimal region ${\mathcal S}(x,dx)$:
\begin{equation}  \label{weightdef}
\hat{\sigma}(x) dx_1 \cdots dx_{n-1} = \frac{\int_{{\mathcal C}(x,dx)} \alpha^2 d\vec{\alpha}}{\int_{\mathcal R} \alpha^2 d\vec{\alpha}}.
\end{equation}
Here ${\mathcal C}(x,dx)$ is the cone (within the region ${\mathcal R}$) representing dice
for which the probabilities of the outcomes lie in ${\mathcal S}(x,dx)$:
\begin{equation}
{\mathcal C}(x,dx) = \Big\{\vec{\alpha} \in {\mathcal R}  \Big| x_j \le \frac{\alpha_j^2}{\alpha^2} \le x_j + dx_j\Big\}.
\end{equation}
(Our use of the weighting factor $\alpha^2$ is reminiscent of the ``adjustment'' stage
in the construction of the GAP measure in Refs.~\cite{Goldstein1,Tumulka,Reimann,Goldstein2},
and the integration over ${\mathcal C}(x,dx)$ is reminiscent of the projection stage of that
same construction.)
We can express $\hat{\sigma}(x)$ more formally as
\begin{equation}  \label{formal}
\hat{\sigma}(x) = \frac{\int_{\mathcal R} \left[\prod_{j=1}^{n-1} \delta\Big(x_j - \frac{\alpha_j^2}{\alpha^2}\Big) \right]
\alpha^2 d\vec{\alpha}}{\int_{\mathcal R} \alpha^2 d\vec{\alpha}},
\end{equation}
where $\delta$ is the Dirac delta function.  

{It is not difficult to obtain an explicit expression for $\hat{\sigma}(x)$
starting with Eq.~(\ref{formal}).  For example, in the integral appearing in the numerator of
that equation, one can use the integration variables
$s_1, \ldots, s_{n-1}$ and $\alpha$, where $s_j = \alpha_j/\alpha$.  Then $d\vec{\alpha}$
becomes $(1/s_n)\alpha^{n-1} ds_1 \ldots ds_{n-1} d\alpha$, and the integral 
becomes straightforward.  Here, though, we take
a different path to the same answer, 
starting with Eq.~(\ref{weightdef}).  This latter approach turns out to be more parallel to 
our derivation of the Scrooge distribution in the quantum mechanical setting.}

First note that the numerator in Eq.~(\ref{weightdef}) can be written as
\begin{equation}  \label{num}
\int_{{\mathcal C}(x,dx)} \alpha^2 d\vec{\alpha} = \frac{n}{n+2} \alpha_{max}^2 \cdot \hbox{(volume of ${\mathcal C}(x,dx)$)},
\end{equation}
where $\alpha_{max}$ is the largest value of $\alpha$ over all points in ${\mathcal R}$ satisfying
$\alpha_j^2/\alpha^2 = x_j$ for $j = 1, \ldots, n$.  (We can get Eq.~(\ref{num}) by writing $d\vec{\alpha}$
as $k \alpha^{n-1}d\alpha$, with some constant $k$, for the purpose of integrating over the cone.)  We can find the value of
$\alpha_{max}$ by finding the point of intersection between (i) the ellipsoid that defines
the boundary of ${\mathcal R}$, given by
\begin{equation}
\frac{\alpha_1^2}{(n+2){\mathcal M}_1} + \cdots + \frac{\alpha_n^2}{(n+2){\mathcal M}_n} = 1,
\end{equation}
and (ii) the line, parameterized by $\alpha$, defined by the equations
\begin{equation}
\alpha_j = \sqrt{x_j} \alpha, \hspace{4mm} j = 1, \ldots, n.
\end{equation}
The value of $\alpha$ at this intersection point is
\begin{equation}
\alpha_{max} = \sqrt{\frac{n+2}{\frac{x_1}{{\mathcal M}_1} + \cdots + \frac{x_n}{{\mathcal M}_n}}}.
\end{equation}
We can therefore rewrite Eq.~(\ref{num}) as
\begin{equation}  \label{num2}
\int_{{\mathcal C}(x,dx)} \alpha^2 d\vec{\alpha} =  \frac{n}{\frac{x_1}{{\mathcal M}_1} + \cdots + \frac{x_n}{{\mathcal M}_n}}\cdot \hbox{(volume of ${\mathcal C}(x,dx)$)}.
\end{equation}
Meanwhile, it follows from Eq.~(\ref{geocond}) that the denominator in Eq.~(\ref{weightdef}) is
\begin{equation} \label{denom}
\int_{\mathcal R} \alpha^2 d\vec{\alpha} = ({\mathcal M}_1 + \cdots + {\mathcal M}_n) V_{\mathcal R}.
\end{equation}

{Our next step is to 
compare $\hat{\sigma}(x)$ to the analogous distribution $\hat{\tau}(y)$ 
induced by the uniform distribution of the vector $\vec{\beta}$---the same
$\vec{\beta}$ as in Section 4---over its domain ${\mathcal R}'$.  (Recall
that ${\mathcal R}'$ is the positive section 
of a sphere.)}
\begin{equation} \label{weighttau}
\hat{\tau}(y) dy_1 \cdots dy_{n-1} = \frac{\int_{{\mathcal C'}(y,dy)} \beta^2 d\vec{\beta}}{\int_{\mathcal R'} \beta^2 d\vec{\beta}}.
\vspace{1mm}
\end{equation}
Here  
${\mathcal C'}(y,dy)$ is the cone in ${\mathcal R'}$ for which 
$y_j \le (\beta_j/\beta)^2 \le y_j + dy_j$.  
We can immediately write down an explicit expression for $\hat{\tau}(y)$: it is the same 
as the distribution (\ref{newtau}) on the probability simplex induced by the uniform
distribution over the unit sphere in the $n$-dimensional real Hilbert space---the extra radial
dimension represented by $\beta$ has no bearing on the distribution over the 
probability simplex.  Thus,
\begin{equation}  \label{newtauagain}
\hat{\tau}(y) = \frac{\Gamma(n/2)}{\pi^{n/2}}\cdot \frac{1}{\sqrt{y_1 \cdots y_n}}.
\end{equation}

We work out our expression for $\hat{\sigma}(x)$ by finding the factors by which the 
numerator and denominator in Eq.~(\ref{weighttau}) change when we stretch the sphere in $\vec{\beta}$-space
into an ellipsoid in $\vec{\alpha}$-space.  
{In this transformation (in which $\alpha_j = \beta_j\sqrt{{\mathcal M}_j} $), the relation between $y$ (in Eq.~(\ref{weighttau})) and $x$ 
(in Eq.~(\ref{weightdef})) is given by
$y = g(x)$, where $g$ takes the point $(\alpha_1^2/\alpha^2, \ldots, \alpha_{n-1}^2/\alpha^2)$
in the probability simplex to the point $(\beta_1^2/\beta^2, \ldots, \beta_{n-1}^2/\beta^2)$.} 

Essentially, any appearance of 
${\mathcal M}_j$ in our expression (\ref{weightdef}) for $\hat{\sigma}(x)dx_1\ldots dx_{n-1}$ becomes a 1 in Eq.~(\ref{weighttau}).
Thus, according to Eq.~(\ref{num2}), when we transform from $\vec{\beta}$ to $\vec{\alpha}$,
we multiply the numerator in Eq.~(\ref{weighttau}) by
\begin{equation} \label{numfac}
\frac{ \frac{n}{\frac{x_1}{{\mathcal M}_1} + \cdots + \frac{x_n}{{\mathcal M}_n}}\cdot \hbox{(volume of ${\mathcal C}(x,dx)$)}}{ n\cdot \hbox{(volume of ${\mathcal C'}(y,dy)$)}}.
\end{equation}
And according to Eq.~(\ref{denom}), in this same transformation we multiply the denominator
in Eq.~(\ref{weighttau}) by
\begin{equation} \label{denomfac}
\frac{({\mathcal M}_1 + \cdots + {\mathcal M}_n) V_{\mathcal R}}{n V_{\mathcal R'}}.
\end{equation}
For both the transitions ${\mathcal C'}(y,dy) \rightarrow {\mathcal C}(x,dx)$ and
${\mathcal R'} \rightarrow {\mathcal R}$, the volume increases by the factor
$\sqrt{{\mathcal M}_1 \cdots {\mathcal M}_n}$.  So these volume factors cancel.
Inserting the other factors from Eqs.~(\ref{numfac}) and (\ref{denomfac}), we find that
\begin{equation}  \label{sigma2}
\hat{\sigma}(x) = \hat{\tau}(y){\mathcal J}(y/x) \frac{n}{\left( \frac{x_1}{{\mathcal M}_1} + \cdots +
\frac{x_n}{{\mathcal M}_n}\right) \left( {\mathcal M}_1 + \cdots + {\mathcal M}_n \right)},
\end{equation}
where ${\mathcal J}(y/x)$ is the Jacobian of $y$ with respect to $x$.

Let us now write $y$ explicitly in terms of $x$:
\begin{equation} \label{y}
y_j = \frac{\beta_j^2}{\beta^2} = \frac{\frac{\alpha_j^2}{{\mathcal M}_j}}{\frac{\alpha_1^2}{{\mathcal M}_1}+\cdots +\frac{\alpha_n^2}{{\mathcal M}_n} }
=\frac{\frac{x_j}{{\mathcal M}_j}}{\frac{x_1}{{\mathcal M}_1}+\cdots +\frac{x_n}{{\mathcal M}_n} }.
\end{equation}
From this we can get the Jacobian (very much like the one in
Eq.~(\ref{Jac})):
\begin{equation}  \label{Jac2}
{\mathcal J}(y/x) = \frac{1}{{\mathcal M}_1 \cdots {\mathcal M}_n}\cdot \frac{1}{\big( \frac{x_1}{{\mathcal M}_1} + \cdots + \frac{x_n}{{\mathcal M}_n}\big)^n}.
\end{equation}
Inserting the results of Eqs.~(\ref{y}) and (\ref{Jac2}) into Eq.~(\ref{sigma2}), we arrive at
\begin{equation} \label{realScrooge2}
\hat{\sigma}(x) = \frac{n \Gamma(n/2)}{\pi^{n/2}{\mathcal M} \sqrt{{\mathcal M}_1 \cdots {\mathcal M}_n}\sqrt{x_1 \cdots x_n} \big(\frac{x_1}{{\mathcal M}_1} + \cdots + \frac{x_n}{{\mathcal M}_n}\big)^{\frac{n}{2} + 1}},
\end{equation}
where ${\mathcal M} = {\mathcal M}_1 + \cdots + {\mathcal M}_n$.
This is essentially the same as the expression (\ref{realScrooge}) we obtained earlier
as the real-amplitude Scrooge distribution.  We can make the agreement more explicit by
defining the ratios $\lambda_j = {\mathcal M}_j/{\mathcal M}$,
in which case Eq.~(\ref{realScrooge2}) becomes exactly identical to Eq.~(\ref{realScrooge}),
with these $\lambda_j$'s playing the role of the eigenvalues of the density matrix.  

Note that {we see here} an analog of $\rho$ distortion.  The stretching of the sphere
in $\vec{\beta}$-space into an ellipsoid in $\vec{\alpha}$-space is very much like 
$\rho$ distortion, though in place of the notion of a density matrix, we have a uniform
distribution within the sphere or ellipsoid.  

It may seem that our communication set-up, in which Alice sends a die equipped with
a probabilistic self-destruction mechanism, is rather artificial.  
But the mathematics is actually fairly simple and natural.
We are considering a set of Poisson-distributed random variables and are basically
constructing a measure on the set of values of these variables based on 
distinguishability.  (It is the measure derived from the Fisher information metric.)  
That measure then induces a measure on the probability simplex,
which agrees with the real-amplitude Scrooge distribution.  


\section{A classical interpretation of the complex-amplitude Scrooge distribution}
We now show how to modify the above classical communication scenario 
so as to arrive at the original, complex-amplitude Scrooge distribution.

Not surprisingly, we begin by doubling the number of sides of Alice's dice.  
Let the outcomes be labeled $1_a, 1_b, 2_a, 2_b, \ldots, n_a, n_b$.  
The communication scheme is
exactly as it was in Section 4, except that instead of placing an upper bound
on the expectation value of the number of times each individual outcome 
occurs, we group the $j_a$ and $j_b$ outcomes together and place an upper
bound ${\mathcal M}_j$ on the expectation value of the total number of times the two $j$ outcomes occur.  
We do this for each $j = 1, \ldots, n$.  Again we ask Alice and Bob to maximize the
number of distinguishable signals under this constraint, where ``distinguishable'' again means
having a Fisher-distance separation at least $d_{min}$.  

As before, it is easiest to view the problem in $\vec{\alpha}$-space; let us label the 
variables in the space as $\alpha_{ja}$ and $\alpha_{jb}$.  We are now looking 
for the maximum-volume region ${\mathcal R}$ of the positive section of $\vec{\alpha}$-space
satisfying the constraints
\begin{equation}
\frac{1}{V_{\mathcal R}}\int_{\mathcal R} (\alpha_{ja}^2 + \alpha_{jb}^2) d\vec{\alpha}
= {\mathcal M}_j, \hspace{4mm} j = 1, \ldots, n.
\end{equation}
Let us define the variables $\beta_{ja} = \alpha_{ja}\Big/\sqrt{{\mathcal M}_j}$
and $\beta_{jb} = \alpha_{jb}\Big/\sqrt{{\mathcal M}_j}$.  Then the constraints become
\begin{equation}
\frac{1}{V_{\mathcal R'}}\int_{\mathcal R'} (\beta_{ja}^2 + \beta_{jb}^2) d\vec{\beta}
= 1, \hspace{4mm} j = 1, \ldots, n,
\end{equation}
where ${\mathcal R'}$ is the region in $\vec{\beta}$-space corresponding to 
${\mathcal R}$.  Summing these $n$ constraints, we obtain the equation
\begin{equation}
\frac{1}{V_{\mathcal R'}}\int_{\mathcal R'} \beta^2 d\vec{\beta}
= n,
\end{equation}
where $\beta^2 = \sum_{j=1}^n(\beta_{ja}^2 + \beta_{jb}^2)$.  Maximizing
the volume under this constraint again gives us a sphere in $\vec{\beta}$-space,
which becomes an ellipsoid in $\vec{\alpha}$-space (restricted to the positive
section).  

Continuing as before, we find that the induced probability distribution over the 
\hbox{$(2n - 1)$}-dimensional probability simplex associated with a $2n$-sided die
is the analog of Eq.~(\ref{realScrooge2}), the $n$ values ${\mathcal M}_1, \ldots, {\mathcal M}_n$ now being
replaced by the $2n$ values ${\mathcal M}_1/2, {\mathcal M}_1/2, \ldots, {\mathcal M}_n/2,  {\mathcal M}_n/2$. We also use the definition
$\lambda_j = {\mathcal M}_j/{\mathcal M}$.  
\begin{equation}  \label{sigma2n}
 \hat{\sigma}_{ab}({\bf x}) = \frac{n \Gamma(n)}{\pi^n \lambda_1 \cdots \lambda_n\sqrt{x_{1a}x_{1b} \cdots x_{na}x_{nb}} \big(\frac{x_{1a}+x_{1b}}{\lambda_1} + \cdots + \frac{x_{na}+x_{nb}}{\lambda_n}\big)^{n + 1}}.
\end{equation}
Here $x_{ja}$ and $x_{jb}$ are the probabilities of the outcomes $j_a$ and $j_b$, and ${\bf x}$ refers
to the point $(x_{1a}, x_{1b}, \ldots, x_{(n-1)a}, x_{(n-1)b}, x_{na})$ in the $(2n-1)$-dimensional probability simplex.  (The value of $x_{nb}$ is
determined by the requirement that the probabilities sum to unity.)

Finally, we obtain a distribution over the $(n-1)$-dimensional probability simplex by 
ignoring the difference between the outcomes $j_a$ and $j_b$.  We can imagine an
observer who cannot see the $a$ and $b$.  (But note that Alice and Bob {\em can}
see this difference.  Otherwise they could not have used all $2n$ dimensions 
of $\vec{\alpha}$-space in their communication.)  For this ``$ab$-blind'' observer,
the distribution of Eq.~(\ref{sigma2n}) looks like the following distribution over the
$(n-1)$-dimensional probability simplex:
\begin{equation} \label{sigman}
\hat{\sigma}(x) = \int \prod_{j=1}^{n-1} \delta[x_j - (x_{ja}+x_{jb})] \hat{\sigma}_{ab}({\bf x}) dx_{1a}
dx_{1b} \cdots dx_{na}.
\end{equation}
Here $\delta$ is the Dirac delta function and the integral is over the $(2n-1)$-dimensional
probability simplex.  

The integral in Eq.~(\ref{sigman}) is straightforward, and one finds that
\begin{equation} \label{Scrooge2}
\hat{\sigma}(x) = \frac{n!}{\lambda_1 \cdots \lambda_n} \cdot \frac{1}{\big( \frac{x_1}{\lambda_1} + \cdots + \frac{x_n}{\lambda_n}\big)^{n+1}}.
\end{equation}
This is the same as the original Scrooge distribution of Eq.~(\ref{Scrooge}).
The role of the eigenvalues of the density matrix is now played by 
the set of values $\lambda_j = {\mathcal M}_j/({\mathcal M}_1 + \cdots + {\mathcal M}_n)$,
where, again, ${\mathcal M}_j$ is the maximum allowed expectation value of the
number of times the outcomes $j_a$ and $j_b$ occur.

\section{Discussion}

In this paper we have shown how the real-amplitude version of the Scrooge distribution
emerges naturally from a classical communication scenario in which
information is transmitted via the values of several random variables $N_j$.  Essentially,
the real-amplitude Scrooge distribution, regarded as a probability distribution over the
probability simplex, is derived from an underlying distribution based on distinguishability.  
Our analysis includes a transformation
that plays something like the role of a $\rho$ distortion: in place of a density matrix,
what is distorted is a distribution over the space of potential signals.

In order to get the original, complex-amplitude Scrooge
distribution for dimension $n$, we needed to consider a case with twice as many random variables,
grouped into pairs,
and then we imagined an observer for whom only the {\em sum} of the variables within each pair was
observable.  

The reader will probably have noticed that the role played by the concept of {\em information} in 
our classical communication problem
seems to be exactly the opposite of the role it plays in the quantum origin of the 
Scrooge distribution.  In quantum theory, the Scrooge distribution is the distribution over
pure states that, upon measurement, provides an observer the {\em least} possible amount of information.
In contrast, in our classical communication scenario, the Scrooge distribution emerges from a requirement
that Alice convey as much information as possible to Bob.  What is common to both cases is
that the information-based criterion favors a distribution that is highly {\em spread out} over the 
probability simplex.  In the quantum case, a distribution spread out over many non-orthogonal states tends to make it difficult for an observer to gain information about the state.  In the classical case, Alice and Bob
want to spread their signals as widely as possible over the {space of possibilities} in order to 
maximize the number of distinguishable signals.  Thus, though the two scenarios are quite different,
their extremization criteria have similar effects.  

An intriguing aspect of our classical scenario is that the probability simplex is
not itself taken as the domain in which the problem is formulated.  Instead, the 
problem is formulated in terms of the number of times each outcome occurs.  The distribution
over the probability simplex 
is a secondary concept, being derived from a more fundamental 
distribution over the space of the numbers of occurrences of the outcomes.  That is, the values $M_j$
are more fundamental in the problem than the probabilities of the outcomes, which are
defined in terms of the $M_j$'s by the equation $x_j = M_j/M$.  In this specific respect, then, the effort to find a 
classical interpretation of the Scrooge distribution seems to lead us away from the models studied
in Refs.~\cite{W1,W3}, in which the set of frequencies of occurrence of the measurement outcomes
was the only source of information considered.

It is interesting to ask whether this feature of our scenario is necessary in order to
get the Scrooge distribution classically.  To address this question, in the Appendix we consider another
classical communication problem, in which {we impose a separate restriction for
each outcome}
as in Section 4, but now with Alice's signals consisting purely of probabilities (which are estimated
by Bob through the observed frequencies of occurrence).  
For simplicity, we
restrict our attention to the most basic case, in which there are only two possible outcomes---so
Alice's die is now a coin to be tossed---and in which we are aiming just for the real-amplitude
Scrooge distribution as opposed to the complex-amplitude version.  We find that
the resulting probability distribution over the probability simplex is {\em not}
of the same form as the real-amplitude Scrooge distribution.  This result can be
taken as one bit of evidence that it is indeed necessary to go beyond the 
probability simplex, and to work in a space of one additional dimension, in order to obtain the Scrooge distribution classically.  {In this connection, it is worth noting that something very similar
can be seen in the research on {\em subentropy}: it happens that certain simple relations between
subentropy and Shannon entropy can be obtained only by
lifting the normalization restriction that defines the probability simplex and working in the larger space
of unnormalized $n$-tuples \cite{Jozsa1,Jozsa3}.   
}

Finally, one might wonder about the potential significance of our need to invoke an
``$ab$-blind'' observer in order to obtain the complex-amplitude Scrooge distribution. 
It is well known that the number of independent parameters required to specify a pure quantum
state (of a system with a finite-dimensional Hilbert space) is exactly twice the number
of independent probabilities associated with a complete, orthogonal measurement on 
the system.  Here we are seeing another manifestation of this factor of two: the 
classical measurement outcomes, corresponding to the sides of a rolled die, have to be
grouped into pairs, and we need to imagine an observer incapable of distinguishing 
between the elements of any pair.  In our actual quantum world, one can reasonably ask whether there is any interesting sense
in which we ourselves are ``$ab$-blind.''  But this question lies well beyond the scope of the 
present
paper.

\vspace{6pt} 

\conflictsofinterest{The author declares no conflict of interest.} 



\appendixtitles{yes} 
\appendixsections{one} 
\appendix
\section{Communicating through probabilities}
Here we consider a classical communication problem based directly on probabilities,
as opposed to being based on the number of times each outcome occurs.  We restrict our
attention to the case of two outcomes,
which we imagine as ``heads'' and ``tails'' for a tossed coin.  The question is whether the real-amplitude
Scrooge distribution for $n=2$ can be obtained in this way.  

Alice is trying to convey to Bob the identity of a point in the one-dimensional probability
simplex (not the two-dimensional space with axes labeled ``number of heads'' and ``number of tails").
The ``simplex'' in this case is just a line segment, and the points of the simplex are labeled by the probability $x$ of heads. (The probability
of tails is $1 - x$.)  Alice conveys her signal by sending Bob a coin with weights
$(x, 1-x)$.  Bob tosses the coin in order to estimate the value of $x$, but he is allowed
to toss it only $N$ times, at which point the coin will self-destruct.  
 Alice chooses in advance
a set of points in the probability simplex that will serve as her potential signals, and
she provides Bob with the list of these points.  
Alice also chooses a function $N(x)$ that determines how many times Bob will be able to 
toss the coin if the coin's weights are $(x, 1-x)$.  But Bob does not know the function $N(x)$
and
is not allowed to use the observed total number of tosses in his estimation of the value of $x$.  He
can use only the frequencies of occurrence of heads and tails.  

We limit the amount of information Alice can convey per coin by specifying the values of two quantities:
(i) the expectation value ${\mathcal N}$ of the total
number of tosses, and  
(ii) the expectation value ${\mathcal N}_H$ of the number of heads.
If we let $\rho(x)dx$ be the number of signals lying
between the values $x$ and $x + dx$, we can write these two restrictions as follows:
\begin{equation} \label{c1}
\int_0^1 N(x) \rho(x) dx = {\mathcal N} \int_0^1 \rho(x) dx.
\end{equation}
\begin{equation} \label{c2}
\int_0^1 x N(x) \rho(x) dx = {\mathcal N}_H \int_0^1  \rho(x) dx.
\end{equation} 

As before, we insist that Alice choose the signal values so that neighboring
signals have a certain minimum degree of distinguishability as quantified by the 
Fisher information metric.  For the binomial distributions we are dealing with here,
this condition works out to be
\begin{equation}
\Delta x = \sqrt{\frac{x(1-x)}{N(x)}} \, d_{min},
\end{equation}
where $\Delta x$ is the separation between successive signals.  The density
$\rho(x)$ of signals is therefore
\begin{equation} \label{signaldensity}
\rho(x) = \frac{1}{\Delta x} = \sqrt{\frac{N(x)}{x(1- x)}} \, \frac{1}{d_{min}}.
\end{equation}

Alice wants to maximize the number of distinct signals.  So, in choosing the function
$N(x)$, she needs to solve the following optimization problem: maximize the quantity
(from Eq.~(\ref{signaldensity}))
\begin{equation}
\int_0^1 \sqrt{\frac{N(x)}{x (1-x)}} dx,
\end{equation}
while satisfying the following two constraints (which come from Eqs.~(\ref{c1}) and (\ref{c2}),
combined with Eq.~(\ref{signaldensity})):
\begin{equation} \label{constraintone}
\int_0^1 \frac{N(x)^{3/2} - {\mathcal N}N(x)^{1/2}}{\sqrt{x(1-x)}} dx = 0.
\end{equation}
\begin{equation}
\int_0^1 \frac{xN(x)^{3/2} - {\mathcal N}_H N(x)^{1/2}}{\sqrt{x(1-x)}} dx = 0.
\end{equation}
This problem can be solved by the calculus of variations, and one finds
that Alice should choose $N(x)$ to be of the form
\begin{equation} \label{Nequation}
N(x) \propto \frac{1}{\frac{x}{\lambda} + \frac{1-x}{1-\lambda}}.
\end{equation}
Here $\lambda$ is a real number between zero and one,
fixed by the requirement that the overall probability of heads must equal 
${\mathcal N}_H/{\mathcal N}$.  (We could have written the result in other ways;
we use $\lambda$ only to facilitate our later comparison with the Scrooge distribution.)
Once the value of $\lambda$ is set, the constant factor multiplying the right-hand side is fixed by Eq.~(\ref{constraintone}).

We now use this result to generate a probability distribution $\hat{\sigma}(x)$ over the probability 
simplex.  We define it as follows: in many rounds of communication, we want $\hat{\sigma}(x)dx$ to approximate the fraction of the total number of tosses that come from
a coin whose probability of heads is between
$x$ and $x + dx$.  More precisely, we define $\hat{\sigma}(x)$ to be proportional
to $N(x)\rho(x)$, with the 
proportionality constant set by the normalization condition $\int_0^1 \hat{\sigma}(x) dx = 1$.
(We have multiplied $\rho(x)$ by $N(x)$ to turn the density of signals into the density of tosses.) 
Substituting for $N(x)$ and $\rho(x)$ in accordance with Eqs.~(\ref{signaldensity})
and (\ref{Nequation}), we arrive at
\begin{equation}
\hat{\sigma}(x) = \frac{A}{\sqrt{x (1-x)}}\cdot \frac{1}{\big( \frac{x}{\lambda}+\frac{1-x}{1-\lambda}\big)^{3/2}},
\end{equation}
where $A$ is the normalization constant.  Comparing this form with that of Eq.~(\ref{realScrooge}),
we see that this alternative problem does not lead us to the real-amplitude Scrooge
distribution: the exponent appearing in the denominator is $3/2$ instead of 2.  
Moreover, $\lambda$ and $1-\lambda$ have no obvious meaning in this problem, whereas
in the problem considered in Sections 4 and 5, the $\lambda_j$'s can be interpreted directly
in terms of the imposed bounds ${\mathcal M}_j$ on the expectation values of the 
number of times the various outcomes occur.


\reftitle{References}





\end{document}